\documentclass[aps,pra,twocolumn,showpacs,preprintnumbers,amsmath,amssymb,superscriptaddress,longbibliography]{revtex4-1}

\usepackage{mathrsfs}
\usepackage{amsfonts}
\usepackage{amsmath}
\usepackage{amssymb}
\usepackage{natbib}
\usepackage{bbm}
\usepackage{hyperref}
\usepackage{xcolor}

\usepackage{graphicx}

\begin{document}
	
	
\title{Measurement operator for quantum nondemolition measurements}

	
	\author{Ebubechukwu O. Ilo-Okeke}
    \thanks{The indicated authors are joint first authors}
	\affiliation{New York University Shanghai, 1555 Century Ave, Pudong, Shanghai 200122, China}  
	\affiliation{Department of Physics, School of Physical Sciences, Federal University of Technology, P. M. B. 1526, Owerri 460001, Nigeria}

	\author{Ping Chen}
    \thanks{The indicated authors are joint first authors}
	\affiliation{State Key Laboratory of Precision Spectroscopy, School of Physical and Material Sciences, East China Normal University, Shanghai 200062, China}
	\author{Shuang Li}
	\affiliation{State Key Laboratory of Precision Spectroscopy, School of Physical and Material Sciences, East China Normal University, Shanghai 200062, China}
	\author{Bede C. Anusionwu}
	\affiliation{Department of Physics, School of Physical Sciences, Federal University of Technology, P. M. B. 1526, Owerri 460001, Nigeria}
	\author{Valentin Ivannikov}
	\affiliation{New York University Shanghai, 1555 Century Ave, Pudong, Shanghai 200122, China}
	\affiliation{State Key Laboratory of Precision Spectroscopy, School of Physical and Material Sciences,	East China Normal University, Shanghai 200062, China}
	\author{Tim Byrnes}
 \email{tim.byrnes@nyu.edu}
	\affiliation{New York University Shanghai, 1555 Century Ave, Pudong, Shanghai 200122, China}
 	\affiliation{State Key Laboratory of Precision Spectroscopy, School of Physical and Material Sciences, 		East China Normal University, Shanghai 200062, China}
	\affiliation{NYU-ECNU Institute of Physics at NYU Shanghai, 3663 Zhongshan Road North, Shanghai 200062, China}
 \affiliation{Center for Quantum and Topological Systems (CQTS), NYUAD Research Institute, New York University Abu Dhabi, UAE.}
	\affiliation{Department of Physics, New York University, New York, NY 10003, USA}

	\date{\today}
	
	\begin{abstract}
We derive a measurement operator corresponding to a quantum nondemolition (QND) measurement of an atomic ensemble. The quantum measurement operator takes the form of a positive operator valued measure (POVM) and is valid for arbitrary interaction times, initial coherent state amplitudes, and final photon measurement outcomes.  We analyze the dependence on various parameters and show that the effect of the QND measurement for short interaction times is to apply a Gaussian modulation of the initial state wavefunction.  We derive approximate expressions for the POVM in various limits, such as the short interaction time regime and projective measurement limit.  Several examples are shown which shows how spin squeezing and Schrodinger cat states can be generated using the measurement.  
	\end{abstract}

	\maketitle

	\section{Introduction \label{sec:intro}}

Quantum nondemolition (QND) measurements~\cite{braginsky1980,grangier1998} are one of the established tools for measuring and engineering quantum mechanical systems.
Typically, a quantum measurement disturbs the quantum state of a system, due to the backaction of the measurement on the system.  Furthermore, due to the free evolution of the Hamiltonian, repeated measurements do not leave the state invariant, as often considered in a textbook example of projective quantum measurements.  In order to qualify as a QND measurement, the measurement observable and the free Hamiltonian must commute.  In a sequence of measurements, the observables in the sequence must also commute.  In this situation, QND measurements can be considered a type of measurement that least disturbs the system, and were investigated in the context of 
precision measurement applications such as gravitation wave detection \cite{eberle2010,pitkin2011}, squeezed state preparation \cite{scully1997,loudon2000,byrnes2021} in optical \cite{brune1990,holland1991,ueda1992}, and mechanical oscillators \cite{lecocq2015}. 

\begin{figure}[t]
\includegraphics[width=\columnwidth]{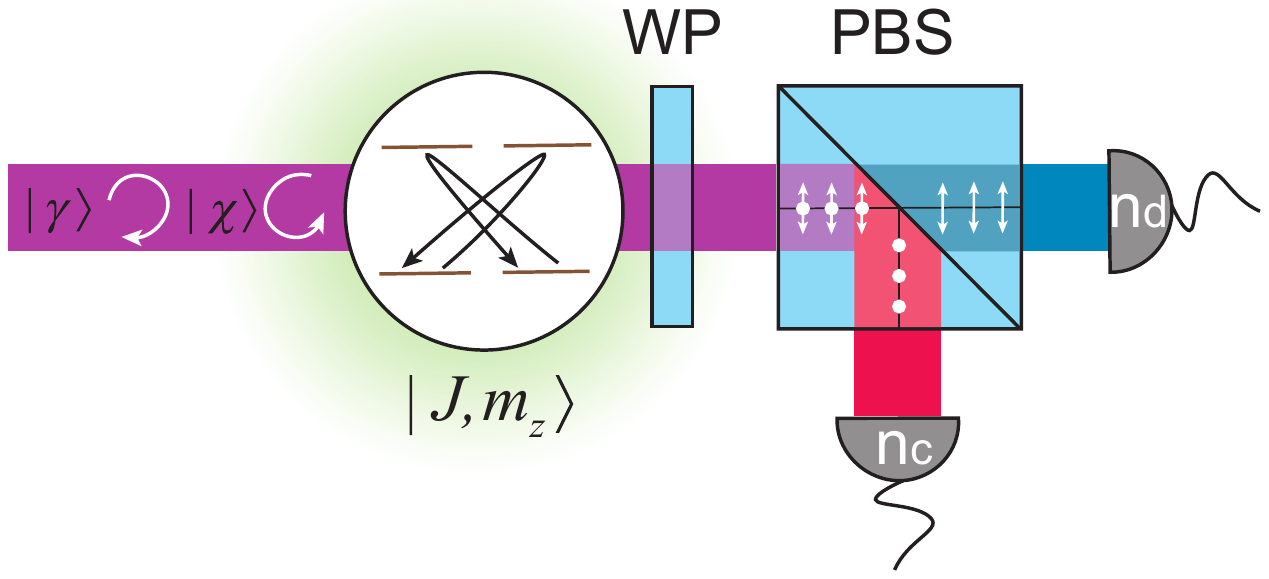}
\caption{Setup for quantum nondemolition measurement of atomic gases considered in this paper. Atoms in the cloud interact with an incident laser beam consisting of two circularly polarized light modes detuned from atomic resonant transition. The spin dependent atom-light interactions cause parts of the laser beam to accumulate a phase relative to the other parts.
The atoms in the atomic cloud consist of ground states of spin-$j$, which produce a state that can be written in terms of the collective spin $ | J, m_z \rangle $ of the cloud. By passing the light beams through a wave plate (WP), the two light polarizations are interfered. 
The polarizing beam splitter (PBS) and photodetectors detects the light in the linear polarization basis, producing photocounts $ n_c $ and $ n_d $. }
\label{fig1}
\end{figure}

In atomic systems, QND measurements are well-known for realizing squeezed states~\cite{takahashi1999,kuzmich2000,higbie2005,meppelink2010,schleier-smith2010,vasilakis2015,moller2017,behbood2014generation,takano2009spin,bao2020spin}. The general principle is shown in Fig. \ref{fig1}.  A superposition of clockwise and anti-clockwise circularly polarized light illuminates an atomic cloud.  The dispersive coupling of the light with the atomic ground states produces a spin-dependent phase that is different for the two types of circularly polarized light.  A waveplate interferes the two light modes such that the spin dependent phase becomes an observable, and a polarizing beam splitter and photodetectors measures the light to detect this.  Due to the spin-dependence of the phase, this amounts to an indirect measurement of the spins of the atoms.  Typically, the interaction between the atoms and the light is weak, such that there is a limited collapse of the atoms' wave function. Such a procedure produces a squeezed state in the spin variables due to the partial collapse of the wavefunction due to measurement. Besides squeezed state preparation, works focusing on non-Gaussian correlated quantum states preparation, such as supersinglets~\cite{cabello2003Review}, Schrodinger cat states, and N00N states,  and preparation of correlated states between spatially separated atomic ensembles~\cite{julsgaard2001} have been proposed. These states are typically highly entangled states that play a central role in various quantum protocols such as entanglement purification~\cite{bennett1996b,bennett1996c}, quantum teleportation~\cite{bennett1996,horodecki1999,pyrkov2014}, remote state preparation~\cite{chaudhary2021}, clock synchronization~\cite{jozsa2000,ilo-okeke2018}, quantum computing~\cite{byrnes2012,abdelrahman2014}.



In this paper, we develop a theory of QND measurements in terms of a positive operator-value measure (POVM).  POVMs are generalized measurement operators that do not necessarily follow the form of a projection operator.  In our case, depending upon the strength of the interactions, the POVM can result in a weak or strong measurement. Our theory is valid for a wide range of regimes, and can be applied to any atom-light interaction time, any initial coherent state amplitude and any photonic measurement outcome.  By writing the QND measurement as a POVM, it allows one to better grasp what is feasible with the technique, in a succinct mathematical form.  We look at particular limits of the theory, such as the short interaction time limit where the effect of the measurement produces a Gaussian modulation of the amplitude of the initial state, written in terms of total spin $ J_z $ eigenstates.  Another limit we examine is the projective limit where the state collapses to a single $J_z $ eigenstate.  We also show that for long atom-light interaction times the measurements have some highly non-Gaussian effects, where rather exotic measurements can be performed, capable of realizing Schrodinger cat states.

This paper is organized as follows. In Sec.~\ref{sec:Dispersiveimaging} we introduce the formalism to define our measurement operator for QND measurements.  In Sec. \ref{sec:ampfunc} we discuss further the amplitude function which plays a central role to modify the nature of the atomic wavefunction.  In Sec.~\ref{sec:ProjectionMecurement} we show how a projective measurement can be obtained in a particular limit of the QND measurement.  We show several examples of the measurement operator to produce various states in Sec. \ref{sec:examples}.  Finally, we summarize and conclude in Sec. \ref{sec:Discussion}.

	\section{POVM for QND measurements}\label{sec:Dispersiveimaging}
	
We consider the experimental configuration as shown in Fig.~\ref{fig1}.  Coherent light consisting of left and right circularly polarized light, which we denote by bosonic annihilation operators as $ a,b $, illuminate the atomic ensemble.  The light is chosen at a frequency which is detuned from atomic resonance transition, producing a second order dispersive interaction.  Within the rotating wave approximation and applying the adiabatic elimination, the effective Hamiltonian governing the evolution is described as~\cite{ilo-okeke2014} 
\begin{align}
		\hat{H} = \hbar g J_z S_z , 
		\label{hamint}
\end{align}
where $g$ is the atom light coupling frequency, $J_z$ is the \emph{z}-component of the total spin operator of the atomic ensemble. The atomic degrees of freedom consist of \emph{N} particles of spin-\emph{j}, $J_\mathrm{max} = Nj$ is the maximum spin angular momentum, and $J$ takes on values $J = 0, \cdots, J_\mathrm{max}$.  The eigenstates are denoted $ J_z | J, m_z \rangle = m_z | J, m_z \rangle$, where $ m_z = -J, -J+1, ... , J $. The spins typically refer to the hyperfine ground states of the atoms.  The light modes are written using the Stokes operator $ S_z = (a^\dagger a - b^\dagger b)/2$.  The form of the interaction (\ref{hamint}) originates from the fact that the two types of circularly polarized light interact differently to the spins of the atoms.  

The initial state of the light that illuminates the atomic ensemble is taken to be in a coherent state of the form
\begin{align}
\lvert \gamma \rangle \lvert \chi\rangle = e^{-|\gamma|^2/2 } e^{-|\chi|^2/2 } e^{\gamma a^\dagger}  e^{\chi b^\dagger} \lvert 0\rangle,
\end{align} 
where $ \chi,\gamma $ are complex amplitude. The optimal case involves choosing $ \chi = \gamma $, which corresponds to linearly polarized light, but we will keep $ \chi,\gamma $ arbitrary for the sake of generality.  The light then interacts with the atoms according to the interaction (\ref{hamint}), which entangles the atoms and the light.  The state at this point becomes 
\begin{align} 
& e^{-i g t J_z S_z} | \gamma \rangle | \chi \rangle | \psi_0 \rangle  = 
| e^{-i g t J_z } \gamma \rangle | e^{i g t J_z} \chi \rangle  | \psi_0 \rangle  \\
& = \sum_{J=0}^{J_{\max}}  \sum_{m_z = - J}^{J}  \langle J, m_z | \psi_0 \rangle 
| e^{-i g t m_z} \gamma \rangle | e^{i g t m_z} \chi \rangle  | J, m_z \rangle ,
\label{halfwaystate}
\end{align}
where we used the fact that $ e^{i \theta a^\dagger a } | \chi \rangle = | e^{i \theta } \chi \rangle  $ for a coherent state. In the second line, we used the completeness identity 
\begin{align}
\sum_{J}\sum_{m_z=-J}^{J}  \lvert J,m_z\rangle\langle J,m_z\rvert = \mathbbm{1}  . 
\label{completeness}
\end{align} 
We note that the terms inside the sum of (\ref{halfwaystate}) have no dependence on $ J $.  Hence the measurement operator acts on all spin sectors and affects the superposition in $ m_z $ only.  


After the light interacts with the atoms, the light fields carry the information about the atoms in their phase. The information is accessed by interfering the modes $ a, b$ with each other using a waveplate according to the following transformation
	\begin{equation}
		\label{eq:BeamSplitter}
		\hat{a}^\dagger = \frac{\hat{c}^\dagger + i\hat{d}^\dagger}{\sqrt{2}},\qquad \hat{b}^\dagger = \frac{+i\hat{c}^\dagger + \hat{d}^\dagger}{\sqrt{2}}.
	\end{equation}
The state after the transformation becomes
\begin{align}
e^{-( \lvert \chi\rvert^2 + \lvert\gamma\rvert^2 )/2}
\exp \left[ \left( \frac{ e^{-i g t J_z/2 } \gamma + i e^{i g t J_z/2} \chi}{\sqrt{2}} \right) c^\dagger  \right]  \nonumber \\
\times \exp \left[ \left(  \frac{ i e^{-i g t J_z/2 } \gamma +  e^{i g t J_z/2} \chi}{\sqrt{2}} \right) d^\dagger  \right]  | 0\rangle | \psi_0 \rangle 
\end{align}
At the final step, the photon numbers in the modes $ c, d $ are detected, yielding $n_c$ and $n_d$ photons respectively.  The unnormalized state after the measurement is 
\begin{align}
		\label{eq:MeasuredState}
		\lvert \psi_{n_c.n_d}\rangle & =\frac{ 
		e^{-( \lvert \chi\rvert^2 + \lvert\gamma\rvert^2 )/2}		}{\sqrt{n_c!n_d!}} 
\left(\frac{\gamma e^{-igt J_z/2  } + i\chi e^{igt J_z/2  }  }{\sqrt{2}}\right)^{n_c}\nonumber\\
		&\times \left(\frac{i\gamma e^{-igt  J_z/2  } + \chi e^{igt  J_z/2 } }{\sqrt{2}}\right)^{n_d} 
		 \lvert\psi_{0}\rangle.
\end{align}
On the right hand side of (\ref{eq:MeasuredState}), we see that the photon operators have disappeared, leaving only the atomic operators. 

Clearly the state $\lvert\psi_{n_c,n_d}\rangle$ can be obtained by the action of the operator $\hat{M}_{n_c,n_d}$ acting on the initial state $\lvert\psi_0\rangle$, $\lvert\psi_{n_c,n_d}\rangle = \hat{M}_{n_c,n_d}\lvert\psi_0\rangle$, where the operator $\hat{M}_{n_c,n_d}$ is defined as~\cite{ilo-okeke2016}  
\begin{align}
\label{eq:measurementoperator}
\hat{M}_{n_c,n_d} = & \frac{ 
		e^{-( \lvert \chi\rvert^2 + \lvert\gamma\rvert^2 )/2}		}{\sqrt{n_c!n_d!}} 
\left(\frac{\gamma e^{-igt J_z/2  } + i\chi e^{igt J_z/2  }  }{\sqrt{2}}\right)^{n_c}\nonumber\\
&\times \left(\frac{i\gamma e^{-igt  J_z/2  } + \chi e^{igt  J_z/2 } }{\sqrt{2}}\right)^{n_d} 
\end{align}
and $n_c$ and $n_d$ are the number of photons detected at detectors $c$ and $d$, respectively.  
The above measurement operator satisfies the decomposition of unity
\begin{equation}
	\label{eq:decomposeunity}
	\sum_{n_c,n_d} \hat{M}^\dagger_{n_c,n_d}\hat{M}_{n_c,n_d} = \mathbbm{1},
\end{equation}
as a consequence of the photon probability being normalized.
	
We now use the fact that a function of a operator $ \hat{A} $ can be decomposed into function of its eigenstates $ \lvert a_n\rangle $ with eigenvalue $ a_n $ %
\begin{align}
		\label{eq:OperatorEqaution}
		f(\hat{A})= \sum_{n} f(a_n) \lvert a_n\rangle\langle a_n \rvert . 
\end{align}
We may hence write the measurement operator (\ref{eq:measurementoperator}) as
%
\begin{equation}
	 \label{eq:measurementoperator_spectral}
	\begin{split}
		&\hat{M}_{n_c,n_d}  = e^{-( \lvert \chi\rvert^2 + \lvert\gamma\rvert^2 )/2}  \left(\frac{\lvert \gamma\rvert^2 + \lvert\chi\rvert^2}{2} \right)^{\frac{n_c + n_d}{2}}   \\
			& \times \sum_{J=0}^{J_\mathrm{max}}\sum_{m_z=-J}^{J}   e^{i(n_c \phi_{c}(m_z) + n_d \phi_{d}(m_z) )}  \\
			&\times A_{n_c n_d} (m_z)  \lvert J,m_z\rangle\langle J, m_z\rvert,
		\end{split}
\end{equation}
where we defined the amplitude
\begin{align}
A_{n_c n_d} (m_z) & =   \frac{\left( 1 + \cos2\eta\cos2\phi(m_z) \right)^{\frac{n_c}{2}}}{\sqrt{n_c!}}\nonumber \\
		& \times \frac{\left(1 - \cos2\eta \cos2\phi(m_z) \right)^{\frac{n_d}{2}}}{\sqrt{n_d!}}
  \label{afuncdef}
\end{align}
phases
\begin{equation}
		\label{eq:phases}
		\begin{split}
\phi(m_z)  & = \frac{gt m_z}{2} + \frac{\phi_{\chi\gamma}}{2} + \frac{\pi}{4}  ,\\
\phi_{c} (m_z) & = \arctan\left(\tan\eta \tan\phi(m_z) \right),\\
\phi_{d} (m_z) & = \arctan\left( \cot \eta  \tan\phi (m_z) \right) - \frac{\pi}{2}, \\
\phi_{\chi\gamma} & = \text{arg} (\chi) - \text{arg} (\gamma)
		\end{split}
\end{equation}
and we have removed irrelevant global phases. Here we defined $\eta$ as the ratio 
	\begin{equation}
		\label{eq:lightamplitude}
		\tan\eta = \frac{\lvert\chi\rvert - \lvert\gamma\rvert}{\lvert\chi\rvert + \lvert\gamma\rvert}.
	\end{equation}
In (\ref{eq:measurementoperator_spectral}) we used the completeness relation. 

We note that in the case that $ | \chi | = | \gamma |$, $ \eta = 0 $ and the $m_z $ dependence on $ \phi_{c} (m_z), \phi_{d} (m_z)$ disappears. The amplitude function (\ref{afuncdef}) is purely real and has the effect of modulating the wavefunction.  

The probability of a particular photonic measurement outcome is given by
\begin{align}
p_{n_c n_d} = & \langle \psi_0 | \hat{M}_{n_c,n_d}^\dagger \hat{M}_{n_c,n_d}  | \psi_0 \rangle \nonumber \\
 = & e^{-( \lvert \chi\rvert^2 + \lvert\gamma\rvert^2 )}  \left(\frac{ \lvert \gamma\rvert^2 + \lvert\chi\rvert^2}{2} \right)^{n_c + n_d } \nonumber \\
 & \times \sum_{J=0}^{J_\mathrm{max}}\sum_{m_z=-J}^{J} | A_{n_c n_d} (m_z)  \langle J, m_z\rvert \psi_0 \rangle  |^2   .
 \label{probformula}
\end{align}

\section{The amplitude function}
\label{sec:ampfunc}

The final form of the POVM for QND measurements (\ref{eq:measurementoperator_spectral}), while exact, is not yet in a form that is easily grasped intuitively. In this section we give a more intuitive description of the effect of the POVM. Specifically we study the form of the amplitude function $ A_{n_c n_d} (m_z) $.

\subsection{Parameter dependence}

\begin{figure}[t]
\includegraphics[width=\columnwidth]{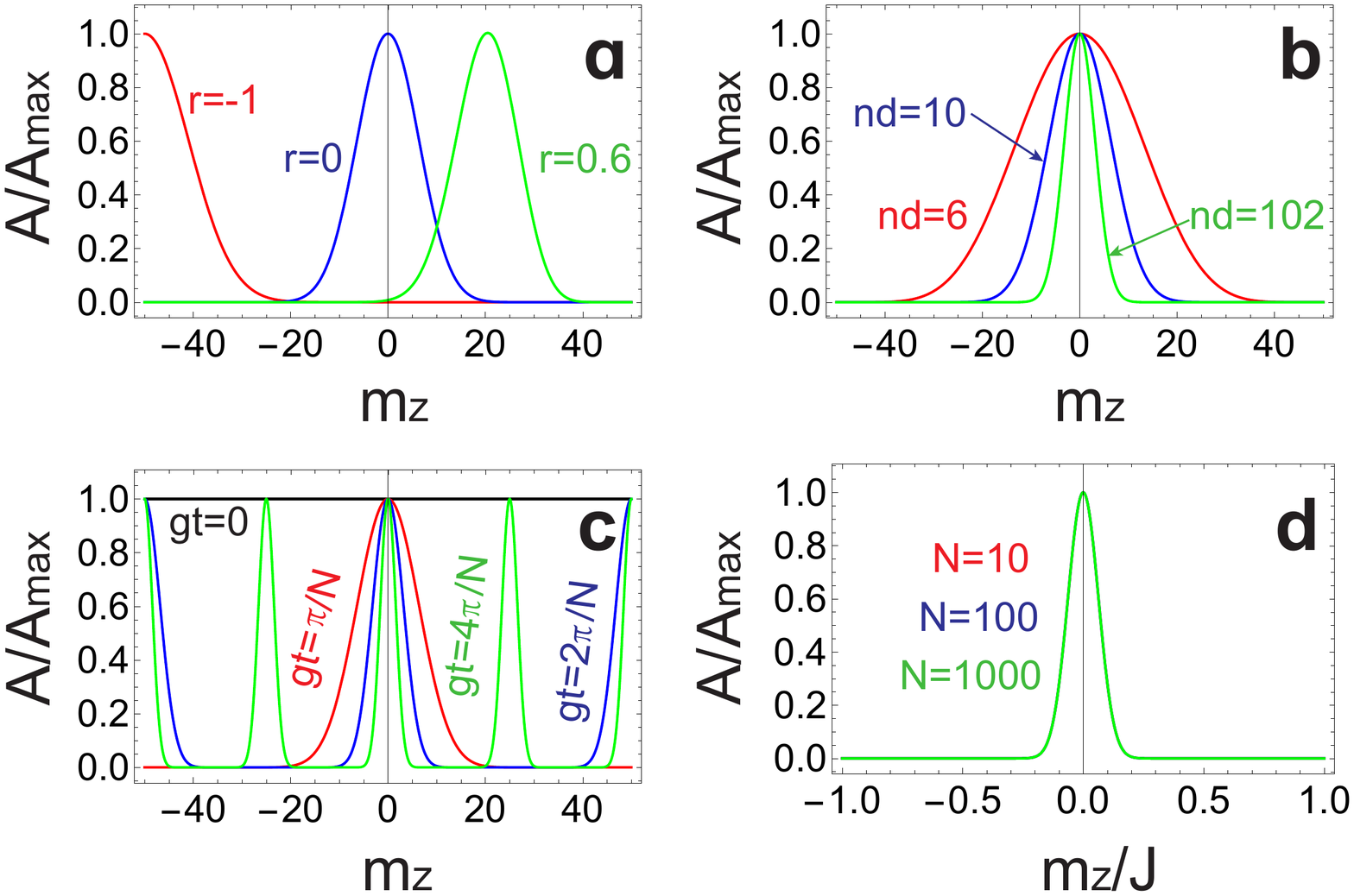}
\caption{The amplitude function $ A_{n_c n_d} (m_z) $ as defined in (\ref{afuncdef}). Plotted are the amplitude functions as a function of $ m_z $ normalized to their maximum values.  (a) Varying the ratio (\ref{normratio}) for interaction time $ gt = \pi/ N$.  (b) Varying the total photon number while keeping $ r $ fixed for  $ gt = \pi/ N$.  (c) Varying the interaction time $ g t $ while for $ n_c = n_d = 25 $.  (d) Varying $ N $ for interaction times $ gt = \pi/ N$ keeping $ n_c = n_d = 25 $ fixed. Common parameters are $ \gamma = 5.1, \chi = 5 $ with the exception of (b) where $ \gamma = 2.55, \chi = 2.5 $ was chosen for the $ n_c = n_d = 6 $ case and  $ \gamma = 10.2, \chi = 10 $ was chosen for the $ n_c = n_d = 102 $ case. }
\label{fig2}
\end{figure}

We first study the amplitude function  $ A_{n_c n_d} (m_z) $ numerically, and capture its basic behavior as a function of various parameters.  Figure \ref{fig2} shows the functional dependence of the amplitude function with $ m_z $, while varying various parameters. 

In Fig.  \ref{fig2}(a), we show the effect of different ratios of the measurement outcome
\begin{align}
r_{n_c n_d}  = \frac{n_d - n_c}{n_c + n_d},
\label{normratio}
\end{align}
keeping $ n_c + n_d $ fixed.  We see that the amplitude function generally has the form of a Gaussian, but is centered at different values of $ m_z $, depending on the value of $ r $. By solving $ d A/dm_z = 0 $ we can find location of the Gaussian to be
\begin{align}
\cos 2 \eta \sin ( gt m_z + \phi_{\chi \gamma}) = r_{n_c n_d}  = \frac{n_d - n_c}{n_c + n_d} .
\label{peakposition}
\end{align}
For a choice of time $ gt = \pi/N $, the factor $ -\pi/2 \le gt m_z \le \pi/2 $ and there is a one-to-one relation between the value of $ r_{n_c n_d} $ and the spin $ m_z $.  The factor of $ \cos 2 \eta $ acts to reduce the amplitude of the sine function such that a smaller range of the $ r_{n_c n_d} $ outcomes occur.  

In Fig. \ref{fig2}(b), we keep $ r_{n_c n_d} = 0 $ by setting $ n_c = n_d $ but vary the total photon count $ n_c + n_d $.  We see that larger total photon numbers tends to narrow the distribution.  A similar narrowing effect is also seen in Fig. \ref{fig2}(c), where the photon counts $ n_c, n_d $ are kept constant, but the interaction time $ g t $ is varied.  We see that longer interaction times tend to produce sharper peaks, looking around $ m_z = 0 $.  However, longer interactions tends to also produce additional peaks, due to the additional solutions of (\ref{peakposition}). Such additional peaks occur when $ gt > \pi/N $ and there is no longer a one-to-one relation between $ m_z $ and $ r_{n_c n_d} $.  

Finally, Fig. \ref{fig2}(d) we show the effect of varying the ensemble size $ N $.  In this case, we rescale the horizontal axis so that $ -1 \le m/J \le 1 $, where $ J = N/2$.  The interaction time is set to $ gt = \pi/N $.  We see that the amplitude function under these rescalings are visually identical.  Thus we consider $ gt = \pi/N $ to be the equivalent time regardless of the ensemble size since it affects the spins in an comparable way.

\subsection{Analytical approximation}

The plots of Fig. \ref{fig2} reveal that the amplitude function takes the form of a Gaussian as long as the interaction time is in the range $0< gt \le \pi/N$. We now make an analytical approximation of the amplitude function in this regime, assuming $n_c,n_d\gg1$ such that $ | r_{n_c n_d}  | \ll \cos 2 \eta $.  The amplitude function can be approximated in this regime as
\begin{equation}
\begin{split}
A_{n_c n_d} (m_z) & \approx  \frac{  \left(\frac{2e }{n_c + n_d}\right)^{\frac{n_c+n_d }{2}} }{\left(4\pi^2 n_c n_d \right)^{\frac{1}{4}}} e^{-\frac{1}{2\sigma^2}\left(m_z - m_0\right)^2},
\end{split}
\label{eq:amp_measurementoperator}
\end{equation}
where we have used the Stirling's approximation to rewrite the factorials. This is a Gaussian with variance 
\begin{equation}
		\label{eq:width}
		\sigma^2 = \left( \frac{g^2t^2}{8} \frac{n_c + n_d}{n_cn_d}\left[(n_c + n_d)^2\cos^22\eta - (n_c-n_d)^2\right]\right)^{-1},
\end{equation}
and maximum position at 
\begin{equation}
		\label{eq:peakposition}
		m_0 = \frac{1}{gt}\left[\arcsin\left(\frac{1}{\cos2\eta} \frac{n_d - n_c}{n_d + n_c}\right) - \phi_{\chi\gamma} \right].
\end{equation}
This is the same equation as (\ref{peakposition}) except that only the principal value of the arcsine has been taken. 

Let us examine the form of the formula for the variance and compare it to the results of Fig. \ref{fig2}.  For the case $ | \chi | = | \gamma | $, the variance simplies to 
\begin{align}
\sigma^2 = \frac{2}{g^2 t^2 (n_c + n_d)} \hspace{1cm} (\eta = 0 ) .
\label{sigmasimple}
\end{align}
For this case we can easily see that the variance reduces for longer interaction times and larger photon number.  We also see that choosing a time $ gt = \pi/N $ makes the width $ \sigma \propto N $.  Since the spin variable $ J_z \propto N $, the $ N $-dependence cancels out as was seen in Fig. \ref{fig2}(d).


Substituting (\ref{eq:amp_measurementoperator}) in (\ref{eq:measurementoperator_spectral}) gives the approximated form of the POVM
\begin{equation}
		\label{eq:measurementoperator_approx}
		\begin{split}
& \hat{M}_{n_c,n_d}  = \frac{e^{\frac{n_c + n_d - (\lvert\gamma\rvert^2 + \lvert \chi\rvert^2 ) }{2}}}{\left(4\pi^2 n_c n_d \right)^{\frac{1}{4}}} \left({ \frac{\lvert\gamma\rvert^2 + \lvert\chi\rvert^2}{n_c + n_d}} \right)^{\frac{n_c + n_d}{2}} \\ 
			&\times \sum_{J=0}^{J_\mathrm{max}}\sum_{m_z=-J}^{J} 
			 e^{i(n_c \phi_{c} + n_d\phi_{d})}   e^{-\frac{1}{2\sigma^2}\left(m_z - m_0\right)^2}\lvert J,m_z\rangle\langle J, m_z\rvert.
		\end{split}
\end{equation}
Due to the conservation of photon number during the sequence, the most likely total photon number $n_c + n_d$ to be detected is equal to $\lvert\gamma\rvert^2 + \lvert\chi\rvert^2$.  This may be seen by examining the factors outside the sums in (\ref{eq:measurementoperator_approx}).

	%
	%

The width $\sigma$ offers a way to characterize the measurement operator $\hat{M}$.  For $\sigma$ being large in comparison to the spacing $ \Delta m_z = 1 $ between the eigenvalues of $J_z$, several eigenvalues $| J, m_z \rangle $ would be contained in within its width. Hence a measurement with such width $\sigma > 1 $ would modify the amplitude of the initial state, by applying an Gaussian envelope function centered around $m_0$. Such measurement is often called weak measurement since it gives some information of the eigenvalue spectrum of $J_z$, but not all. According to the trade-off relation between the strength of the measurement and the backaction \cite{ilo-okeke2016}, this also means that the effect on the state is small. As a result, weak measurement would preserve the linear combination of states in the states $ | J, m_z \rangle $ within the width of the measurement operator and does not lead to a total collapse of the quantum state. 

In the opposite limit where the width is small $\sigma < 1 $  in comparison to the spacing  $\Delta m_z = 1 $, only a handful of eigenvalues (and in the very extreme case only one eigenvalue) is contained in the width of the measurement operator $\hat{M}$, and would be centred around $m_0$. As a result, any measurement performed with such width $\sigma \ll 1 $ would produce only one eigenvalue $|J_z,m_z \rangle $ near the peak value $m_0$. 
This is a destructive measurement and leads to the collapse of the quantum state of the system.

	\section{Projection Measurement Limit}
 \label{sec:ProjectionMecurement}

In the previous section we saw that by adjusting the width of the amplitude function one could tune the  measurement operator from a  weak measurement to a projective measurement,   where there is a total collapse of the quantum state.  In this section, we explicitly derive the form of the projection operator, in the limit of a very narrow width of the amplitude function. 

Our starting point to derive the projective limit is (\ref{eq:measurementoperator_approx}). Making the change of variables 
\begin{equation}
		\label{eq:relativebasis}
		\begin{split}
			u & = \frac{n_d + n_c}{2},\\
			v & = \frac{n_d-n_c}{2}.
		\end{split}
\end{equation}
and recognizing that the dominant contribution to the amplitude of the POVM comes from points around $u = (\lvert \gamma\rvert^2 + \lvert\chi\rvert^2)/2$, we write 
	\begin{equation}
		\label{eq:OperatorRelativeBasis}
		\begin{split}
			\hat{M}(u,v)& = \frac{e^{iv\left(\phi_{d,0} - \phi_{c,0} \right)}}{\left[4\pi^2(u^2 -v^2)\right]^{\frac{1}{4}}} e^{-\frac{\left(u - \frac{\lvert \chi\rvert^2 + \lvert\gamma\rvert^2}{2}\right)^2}{\lvert \chi\rvert^2 + \lvert\gamma\rvert^2}}\\
			&\times e^{iu\left( \phi_{c,0} + \phi_{d,0} \right)} \sum_{J=0}^{J_\mathrm{max}}\sum_{m_z=-J}^{J}  
e^{-\frac{1}{2\sigma^2 }\left(m_z - m_0 \right)^2 } \\
& \times e^{-igt\left(m_z - m_0 \right) \left( u \xi_+ -v \xi_- \right)}  \lvert J,m_z\rangle\langle J, m_z\rvert,
		\end{split}
	\end{equation}
where $\phi_{c,0} = \phi_{c}(m_0 )$,  $\phi_{d,0} = \phi_{d}(m_0)$,  and 
\begin{equation}
		\label{eq:PhaseFirstDerivative}
		\begin{split}
  \xi_+ & = 1- \xi_d - \xi_c  \\
  \xi_- & = \xi_d ( m_0) - \xi_c (m_0 ) \\
\xi_c & = \frac{1}{2}\frac{\tan\eta}{\cos^2(\phi(m_0))} \frac{1}{1 +\tan^2\eta\tan^2(\phi(m_0)) },\\
\xi_d & = \frac{1}{2}\frac{\tan\eta}{\cos^2(\phi(m_0))} \frac{1}{\tan^2(\phi(m_0)) + \tan^2\eta}.
		\end{split}
	\end{equation}
In writing (\ref{eq:OperatorRelativeBasis}), we have expanded the phases $\phi_{c}$ and $\phi_{d}$ around $m_0$ assuming they vary linearly in the neighborhood of $m_0$. 

The operator $\hat{M}(u,v)$ satisfies the decomposition of unity, $\sum_{n_c,n_d} \hat{M}^\dagger_{nc,n_d}\hat{M}_{n_c,n_d}= \mathbbm{1} = \int du dv \hat{M}^\dagger (u,v) \hat{M}  (u,v)$,
\begin{align}
    \mathbbm{1} = & \sum_{J,m_z} \lvert J,m_z\rangle\langle J,m_z\rvert \int 2 du\,dv  
    \frac{e^{-2\frac{\left(u - \frac{\lvert \chi\rvert^2 + \lvert\gamma\rvert^2}{2}\right)^2}{\lvert \chi\rvert^2 + \lvert\gamma\rvert^2}}}{\sqrt{4\pi^2(u^2 -v^2)}} \nonumber\\
    &\times e^{-\frac{(m_z - m_0)^2}{\sigma^2}},\nonumber\\
    =& \sum_{J,m_z} \lvert J,m_z\rangle\langle J,m_z\rvert \int du \frac{e^{-2\frac{\left(u - \frac{\lvert \chi\rvert^2 + \lvert\gamma\rvert^2}{2}\right)^2}{\lvert \chi\rvert^2 + \lvert\gamma\rvert^2}}}{\sqrt{\pi u}},\nonumber\\
    =&\sum_{J,m_z} \lvert J,m_z\rangle\langle J,m_z\rvert.
\end{align}
where the factor of $2$ in the numerator is the Jacobian of transformation from $n_c$, $n_d$ to $u$, $v$, and we have used (\ref{eq:peakposition}) in the evaluation of the integral over $v$, $dv = gt\, u\cos2\eta \cos(gt m_0) dm_0 $ in the first line. In the second line, the $u$ in the denominator is evaluated at $(\lvert\chi\rvert^2 + \lvert\gamma\rvert^2)/2$,  while the identity (the last line) simply follows by the definition of completeness relation. Hence we define a projection operator for the operator $\hat{M}(u,v)$ in this limit, $\sigma\ll1$, as 
\begin{equation}
    \label{eq:OperatorRelativeBasis2}
		\begin{split}
            \hat{M} (u,v) =& \frac{e^{-\frac{\left(u - \frac{\lvert \chi\rvert^2 + \lvert\gamma\rvert^2}{2}\right)^2}{\lvert \chi\rvert^2 + \lvert\gamma\rvert^2}}} {\left[\pi g^2t^2 u (u^2\cos^2{2\eta} -v^2)\right]^{\frac{1}{4}}} \\
            &\times\sum_{J=\text{int} (\lvert m_0\rvert ) }^{J_\mathrm{max}}
            \lvert J, \text{int} ( m_0 )  \rangle  \langle J, \text{int} (  m_0 )  \rvert,
		\end{split}
\end{equation}
where $m_0$ is related to $v$ via (\ref{eq:peakposition}), and $ \text{int} ( x )$ rounds $ x $ to the nearest integer. A factor of  $(\pi\sigma^2)^{1/4}$ was introduced to ensure the normalization of the projector $\hat{M} (u,v)$, and we have dropped all irrelevant global phases.

Finally, the form of the (\ref{eq:OperatorRelativeBasis2}) suggests that the projection operator  can be written directly in terms of $u$ and $m_0$, if one multiplies it with the square root of the Jacobian $ gt\, u\cos2\eta \cos(gt m_0)  $ of transformation from $v$ to $m_0$, as
 %
	\begin{equation}
		\label{eq:ProjetionOperatorDirect}
		\begin{split}
            \hat{M} (u,m_0) & =  \frac{e^{-\frac{\left(u - \frac{\lvert \chi\rvert^2 + \lvert\gamma\rvert^2}{2}\right)^2}{\lvert \chi\rvert^2 + \lvert\gamma\rvert^2}} }{\sqrt[4]{\pi u}} \\
            &\times\sum_{J= \mathrm{int}(\lvert m_0\rvert)}^{J_\mathrm{max}}\lvert J, \mathrm{int}( m_0)\rangle\langle J,\mathrm{int}( m_0)\rvert.
		\end{split}
	\end{equation}
The  operator  $\hat{M} (u,m_0) $ consists of product of two parts, the classical part (the measurement current amplitude) that is read of on the measuring device and the projector that performs quantum operation on a quantum state of a system. However, the quantum operator part --- the projection operator --- depends on the classical properties of the measurement device to define it and its action on the quantum state via (\ref{eq:peakposition}). 

\section{Examples}
\label{sec:examples}

Here we show some examples of the formalism introduced in the previous sections where we apply the QND measurements to produce various types of states. These will serve to illustrate several properties of the QND measurements. Here we will consider the case where an atomic ensemble consists of $ N $ atoms occupying two hyperfine ground states such that the effective spin of each atom can be considered to be $ j = 1/2 $.  A typical choice for the two levels may be $ F = 1, m_F = -1 $ and $ F= 2, m_F = 1 $ for $^{87} \text{Rb} $.

\subsection{Spin squeezed states}

Let us assume the initial state of the atoms are prepared in a spin coherent state and given by \cite{byrnes2021}
\begin{align}
|\psi_0 \rangle & = | \theta \rangle \rangle =  \left( \cos \frac{\theta}{2} | \uparrow \rangle + 
\sin \frac{\theta}{2} | \downarrow \rangle  \right)^{\otimes N} \nonumber \\
& = \sum^{J}_{m_z =-J} \sqrt{\binom{N}{  N/2 + m_z}} \nonumber \\
& \times \cos^{ N/2 +m_z } \frac{\theta}{2}  \sin^{N/2-m_z} \frac{\theta}{2}  | J, m_z \rangle ,
\label{spincoherent}
\end{align}
where we denoted the two states of each atom by $ | \uparrow \rangle$, $| \downarrow \rangle $. 
This state has average values of the total spin %
\begin{align}
    \langle J_x \rangle & = \frac{N}{2} \sin \theta    \nonumber \\
    \langle J_z \rangle & = \frac{N}{2} \cos \theta  ,
\end{align}
and normalized variance 
\begin{align}
 \sigma^2_x \equiv \frac{\langle J_z^2 \rangle - \langle J_z \rangle^2}{N^2} = \frac{\sin \theta }{4N} .
\end{align}
The $ \sigma_x \propto 1/\sqrt{N} $ shows that a spin coherent state is in the standard quantum limit.  For an angle choice $ \theta = \pi/2$, the spin coherent state is an eigenstate of $ J_x $ 
\begin{align}
J_x | \theta = \pi/2 \rangle \rangle = \frac{N}{2} | \theta = \pi/2 \rangle \rangle .  
\end{align}

Applying the measurement operator to the spin coherent state (\ref{spincoherent}), we obtain the unnormalized state
\begin{align}
& \hat{M}_{n_c,n_d} | \psi_0 \rangle   = e^{-( \lvert \chi\rvert^2 + \lvert\gamma\rvert^2 )/2}  \left(\frac{\lvert \gamma\rvert^2 + \lvert\chi\rvert^2}{2} \right)^{\frac{n_c + n_d}{2}}   \\
& \times  \sum_{m_z=-J}^{J}   e^{i(n_c \phi_{c}(m_z) + n_d \phi_{d}(m_z) )}   A_{n_c n_d} (m_z) \nonumber \\
& \sqrt{\binom{N}{  N/2 + m_z}} \cos^{ N/2 +m_z } \frac{\theta}{2}  \sin^{N/2-m_z} \frac{\theta}{2}   \lvert J,m_z\rangle . 
\end{align}
The probability of obtaining a particular photon outcome is
\begin{align}
p_{n_c n_d} &  =  e^{- \lvert \chi\rvert^2 - \lvert\gamma\rvert^2 }  \left(\frac{\lvert \gamma\rvert^2 + \lvert\chi\rvert^2}{2} \right)^{n_c + n_d}  \nonumber \\
& \times  \sum_{m_z=-J}^{J} A^2 (n_c,n_d,m_z) \binom{N}{  N/2 + m_z} \nonumber \\
& \times \cos^{ N +2 m_z } \frac{\theta}{2}  \sin^{N-2 m_z} \frac{\theta}{2} .
\end{align}

\begin{figure}[t]
\includegraphics[width=\columnwidth]{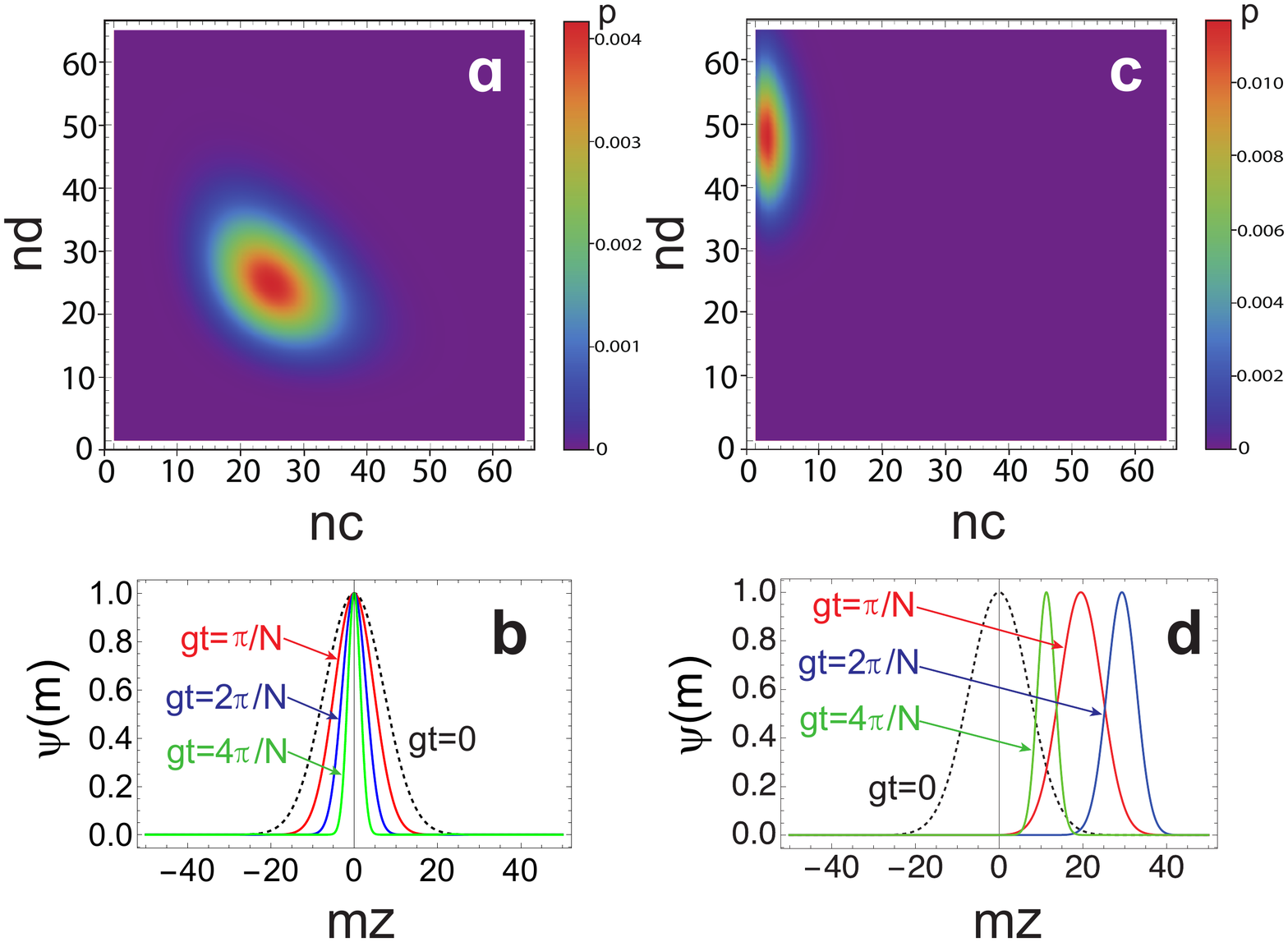}
\caption{Performing a QND measurement on a spin coherent state.  (a)(c) The photon probability $ p_{n_c n_d} $ for and $gt = \pi/N$.  (b)(d) Resulting wavefunction amplitude $ \langle J, m_z | \hat{M}_{n_c,n_d} | \psi_0 \rangle $ for the interaction times as marked.  (a)(b) are results for $ | \psi_0 \rangle = | \theta = \pi/2 \rangle \rangle $;  (c)(d) $ | \psi_0 \rangle = | \theta = \pi/4 \rangle \rangle $. The parameters of the figures are $\chi= 5$, $\gamma = 5.1$, $ N = 100$, $ J = N/2 = 10$.  \label{fig4} }
\end{figure}

In Fig. \ref{fig4}(a)(b) we plot the effect of applying the QND operator to the states $ | \theta = \pi/2 \rangle \rangle$.  First examining the photon outcomes, we see that the most likely outcome occurs when $ n_c = n_d = (|\chi|^2 + | \gamma|^2)/2 $.  Examining different interaction times as shown in Fig. \ref{fig4}(b) we see that the effect of the QND measurement is to create a squeezing effect with respect to $ J_z $, where the original Gaussian distribution of the spin coherent state becomes a narrower distribution.  In Fig. \ref{fig4}(b) the squeezing is  centered around $ m_z =0 $ due to the choice of $ n_c = n_d $, for different choice, the squeezing will be centered around a different $ m_z $ value.  For the longest interaction time $ gt = 4 \pi/N $, additional peaks appear due to the multiple solutions of (\ref{peakposition}).  For current experimental realizations of QND, typically the interaction is in the short time regime, where only a single peak occurs. 

Figure  \ref{fig4}(c)(d) shows the effect of applying the QND operator on the state $ | \theta = \pi/4 \rangle \rangle$.  This corresponds to a state that is polarized along the spin direction $ (J_z + J_x)/\sqrt{2} $, i.e. 45 degrees between the $ z $ and $ x $-axes on the Bloch sphere.  We see that in this case the most likely outcome is $ n_d = |\chi|^2 + | \gamma|^2 $ and $ n_c = 0 $.  Taking this as the typical case, we examine the effect on the wavefunction as shown in Fig. \ref{fig4}(d).  We again see a squeezing effect with increased interaction times.  The location of the peak depends upon the interaction time this time due to the relation (\ref{peakposition}).

\subsection{Dicke states}

Next let us examine the case that the initial state is prepared in an eigenstate of $ J_z $, also known as a Dicke state
\begin{align}
| \psi_0 \rangle = |J, m_z \rangle .
\end{align}
Such a state may be produced in the limit that the QND measurement is a projective measurement, as described in Sec. \ref{sec:ProjectionMecurement}.  While such states are non-trivial to prepare from an experimental perspective, they will illustrate conceptually the notion of QND measurements and show the relationship between the states and photonic outcomes. 

After a QND measurement, the resulting unnormalized state becomes
\begin{align}
\hat{M}_{n_c,n_d} | \psi_0 \rangle  &  = e^{-( \lvert \chi\rvert^2 + \lvert\gamma\rvert^2 )/2}  \left(\frac{\lvert \gamma\rvert^2 + \lvert\chi\rvert^2}{2} \right)^{\frac{n_c + n_d}{2}}   \\
& \times  A_{n_c n_d} (m_z)  \lvert J,m_z\rangle , 
\end{align}
up to a global phase.  Due to the measurement operator (\ref{eq:measurementoperator_spectral}) being diagonal with respect to the spin eigenstates $ | J, m_z \rangle $, any  measurement on any of the spin eigenstates leaves the state unchanged
\begin{align}
\hat{M}_{n_c,n_d} | J, m_z \rangle \propto | J, m_z \rangle .  
\end{align}
We can see the essential effect of the QND measurement where repeated measurements on the spin eigenstates remain unchanged.  More generally, any mixed state consisting of Dicke state will be unchanged
\begin{align}
\hat{M}_{n_c,n_d} \rho_D \hat{M}_{n_c,n_d}^\dagger \propto \rho_D
\end{align}
where 
\begin{align}
\rho_D = \sum_{m_z=-J}^J p(m_z) |J, m_z \rangle 
\langle J, m_z | .  
\end{align}

\begin{figure}[t]
  \includegraphics[width=\columnwidth]{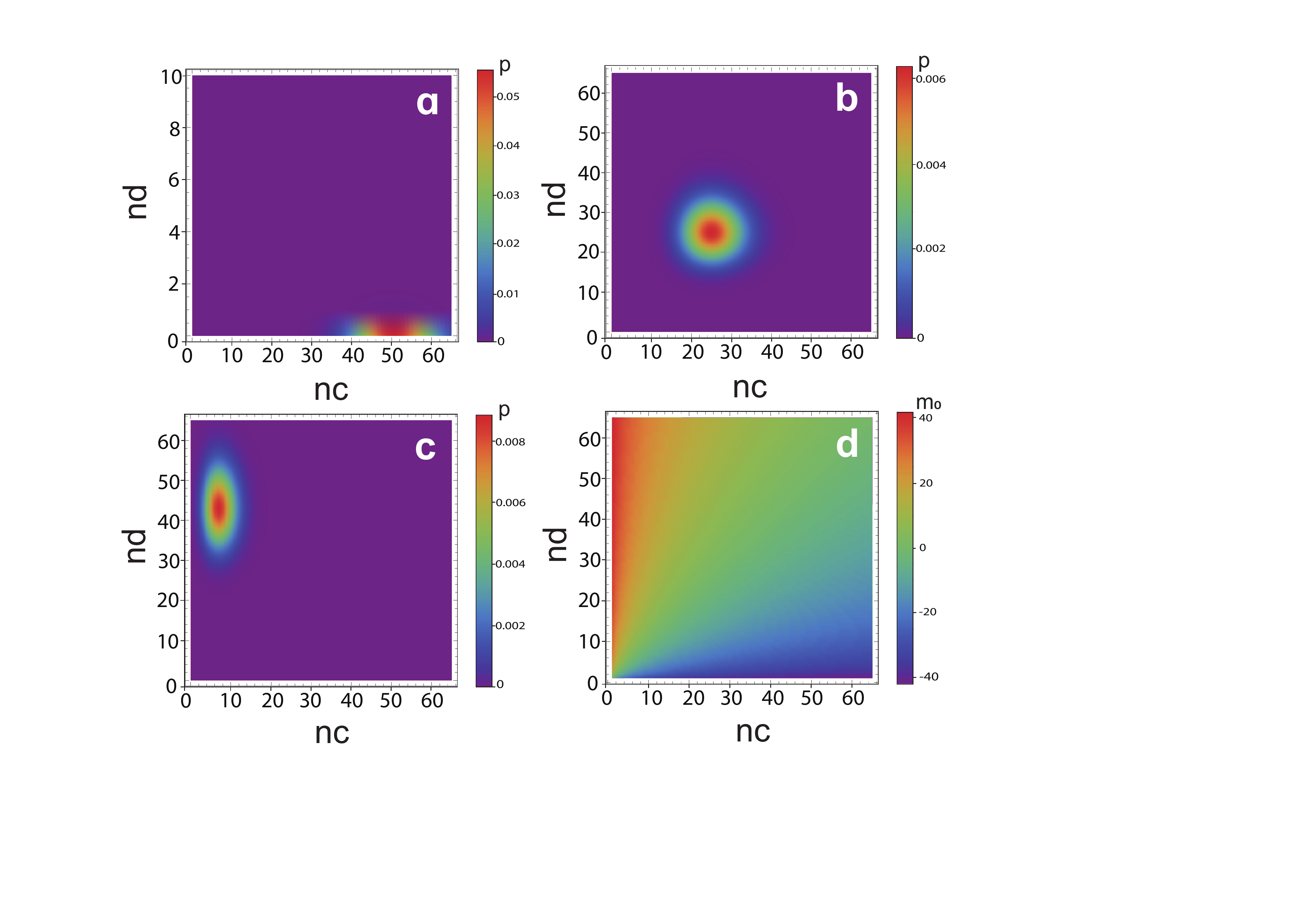}
\caption{The probability density of getting $n_c$ and $n_d$ photons in a measurement of a Dicke state $ | J, m_z \rangle$ with (a)  $m_z=-J$; (b) $m_z = 0 $; (c) $ m_z = J/2$. Map of the relation (\ref{eq:peakposition}) giving the peak value of $ m_0 $ as a function $ n_c, n_d $. The parameters of the figures are $gt = \pi/N$, $\chi= 5$, $\gamma = 5.1$, $ N = 100$, $ J = N/2 = 50$. 
\label{fig3}
}
\end{figure}

The probability of this outcome is
\begin{align}
p_{n_c n_d} &  =  e^{- \lvert \chi\rvert^2 - \lvert\gamma\rvert^2 }  \left(\frac{\lvert \gamma\rvert^2 + \lvert\chi\rvert^2}{2} \right)^{n_c + n_d} A^2 (n_c,n_d,m_z) .
\end{align}
We plot the photonic probability for different values of $ m_z $ in the Dicke state in Fig. \ref{fig3}(a)(b)(c).  We see that depending on the value of $ m_z$, the photonic probabilities emerge with different values of $ n_c, n_d$.  First, the photons emerge along the line 
\begin{align}
n_c + n_d = |\chi |^2 + | \gamma|^2 , 
\end{align}
which is guaranteed by photon number conservation --- the number of photons at the input should be equal to the number on the output. How the photons are distributed along this line, i.e. the value of $ n_c - n_d $, depends upon the value of $ m_z $ according to (\ref{eq:peakposition}).  

The peak value of the photonic distribution in Figs. \ref{fig3}(a)(b)(c) recovers the value of $ m_z $ using the map in Fig. \ref{fig3}(d), which maps the relation (\ref{eq:peakposition}).   We note that due to the finite width of the photonic distribution, at a single shot level, one is not guaranteed to recover the exact $ m_z $ of the initial Dicke state.  This can be attributed to shot noise of the photons which gives fluctuations around the mean.  Hence even in the limit of a completely squeezed Dicke state, the QND measurement gives photonic outcomes which involve noise fluctuations.  We note that for $ m_z = \pm J $, the photonic distributions lie along the $n_c = 0 $ or $ n_d = 0 $ axes respectively, greatly reducing the effect of photonic shot noise (Fig. \ref{fig3}(a)). In these cases, the variance of $ r_{n_c n_d} $ is zero and the value of $ m_z $ can be read even at the single shot level.

\subsection{Cat state generation}

\begin{figure}[t]
  \includegraphics[width=\columnwidth]{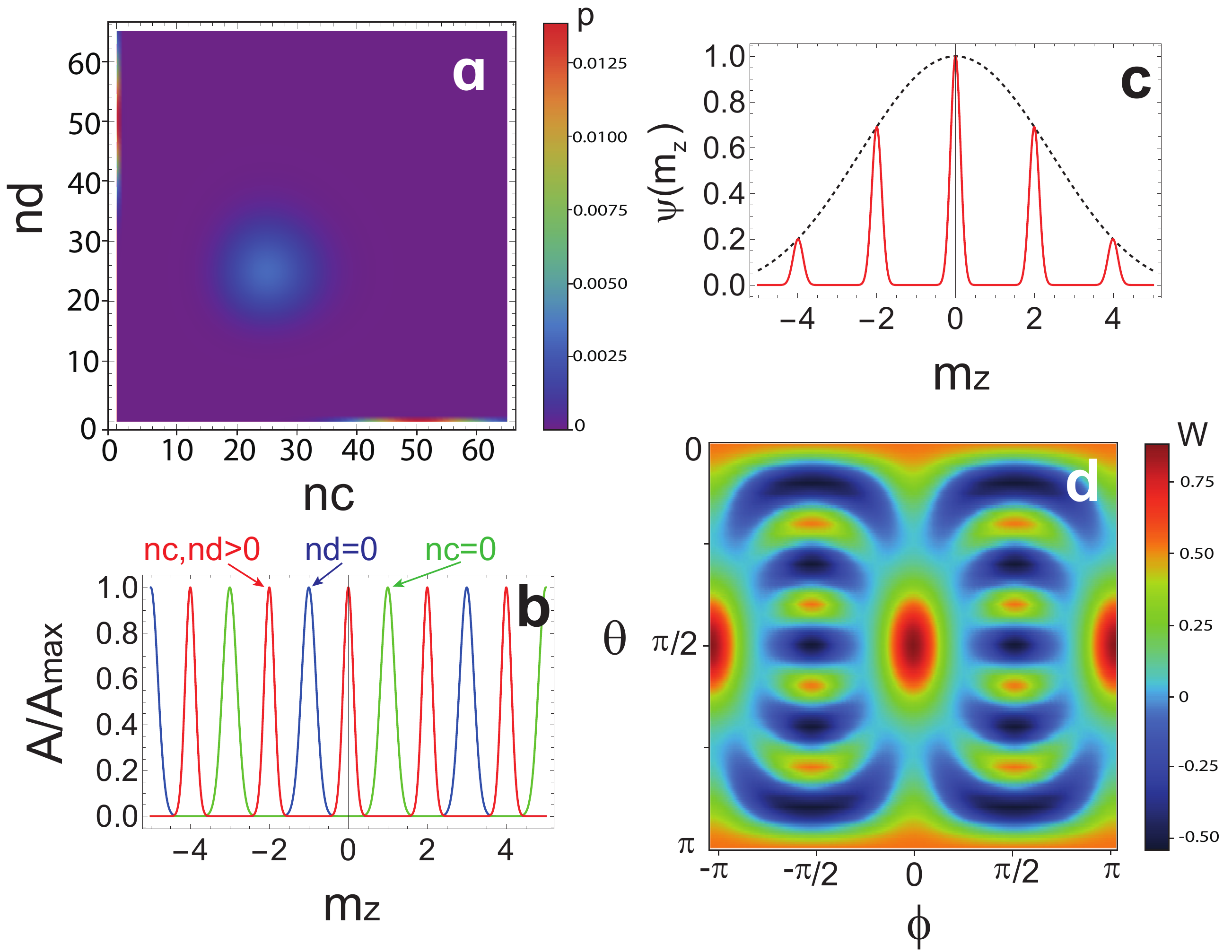}
\caption{Generation of a Schrodinger cat state using QND measurements.  (a) The photonic probability distribution; (b) normalized amplitude function $A_{n_c n_d} (m_z) $; (c) the state before $ \psi (m_z ) =  \langle J, m_z | \psi_0 \rangle = \langle J, m_z | \theta = \pi/2 \rangle \rangle $ (dashed line) and after application of the  QND measurement  $ \psi (m_z ) =  \langle J, m_z |M_{n_c n_d}  | \theta = \pi/2 \rangle \rangle $  (solid line); (d) the Wigner function of the resulting state;   The parameters of the figures are $gt = \pi/2$, $\chi= 5$, $\gamma = 5.1$, $ N = 10$, $ J = N/2 = 5$. 
\label{fig5}
}
\end{figure}

We finally consider the case of long interaction times, where the amplitude function develops non-Gaussian characteristics, as already seen in Fig. \ref{fig2}(c).  Specifically, let us consider interaction times $ gt = \pi/ 2 $, and $ \phi_{\chi \gamma} = 0 $,  $ \eta = 0 $, and even $ N $.  Figure \ref{fig5} shows an example of an application of the QND measurement for the time $ gt = \pi/2 $ to the spin coherent state (\ref{spincoherent}).  In Fig. \ref{fig5}(a) the photon probability is shown, which shows that there are three main outcomes, corresponding to (i) $ n_c =0, n_d \approx |\chi|^2 + | \gamma|^2 $;(ii) $ n_c = n_d \approx (|\chi|^2 + | \gamma|^2)/2 $; and (iii) $ n_d =0, n_c \approx |\chi|^2 + | \gamma|^2 $.  In these three cases the amplitude function simplifies to the following expressions.  For $ n_c>0,  n_d > 0 $  
\begin{align}
A_{n_c n_d} (m_z) = \left\{
\begin{array}{ll}
\frac{2^{(n_c + n_d)/2 }}{\sqrt{n_c! n_d!}}
& \hspace{1cm} m_z \in \text{even} \\
0 & \hspace{1cm} \text{otherwise}
\end{array}
\right.  ,
\label{afunccase1}
\end{align}
for  $ n_c = 0 $, $ n_d \gg 1 $, 
\begin{align}
A_{n_c n_d} (m_z) \approx \left\{
\begin{array}{ll}
\frac{2^{n_d/2 }}{\sqrt{ n_d!}}
& \hspace{1cm} m_z (\text{mod } 4) = 1  \\
0 & \hspace{1cm} \text{otherwise}
\end{array}
\right.  ,
\label{afunccase2}
\end{align}
and for $ n_c \gg 1 $, $ n_d =0 $, 
\begin{align}
A_{n_c n_d} (m_z) \approx \left\{
\begin{array}{ll}
\frac{2^{n_c/2 }}{\sqrt{ n_c!}}
& \hspace{1cm} m_z (\text{mod } 4) = 3  \\
0 & \hspace{1cm} \text{otherwise}
\end{array}
\right.  .
\label{afunccase3}
\end{align}
Figure \ref{fig5}(b) shows the amplitude functions (\ref{afunccase1})-(\ref{afunccase3}).  We see that for these parameters the effect of the QND measurement is to suppress the amplitudes according to the the parity of $ m_z $.

Applying the QND POVM to a spin coherent state for the $ n_c>0,  n_d > 0 $ case, we obtain
\begin{align}
M_{n_c, n_d} | \theta = \pi/2 \rangle \rangle & =
\frac{1}{\sqrt{2^{N/2}}} \sum_{m_z \in \text{even} } \sqrt{\binom{N}{  N/2 + m_z}}  | J, m_z \rangle  \nonumber \\
& =\frac{1}{2} (  | \theta = \pi/2 \rangle \rangle + | \theta = - \pi/2 \rangle \rangle )  
\end{align}
which is a Schrodinger cat state.  Figure \ref{fig5}(c) shows the amplitude of the wavefunction before and after applying the QND measurement. We see that only the even values of $ m_z $ have a non-zero amplitude, as expected from (\ref{afunccase1}).  Figure \ref{fig5}(d) shows the spin Wigner function defined as \cite{dowling1994wigner,byrnes2021} 
\begin{align}
W(\theta, \phi) = \sum_{L=0}^{2J} \sum_{M=-L}^{L} \rho_{LM} Y_{LM} (\theta, \phi), \label{wigner}
\end{align}
where $Y_{LM} (\theta, \phi) $ are the spherical harmonic functions.  Here, $\rho_{LM}$ is defined as 
\begin{align}
\rho_{LM}= &  \sum_{m_z=-J}^{J} \sum_{m_z'=-J}^{J} (-1)^{j-m_z-M} 
\langle J m_z; J -m_z' | L M \rangle \nonumber \\
& \times \langle J m_z | \rho | J m_z' \rangle,
\end{align}
where $ \langle J m_z; J m_z' | L M\rangle  $ is a Clebsch-Gordan coefficient\index{Clebsch-Gordan coefficient} for combining two angular momentum eigenstates $ |J m_z \rangle $ and $ |J m_z' \rangle $ to $ | L M \rangle $, and $ \rho $ is the density matrix.  The Wigner function shows the characteristic interference fringes between the two dominant probability peaks around $ \theta = \pm \pi/2 $. This indicates the highly non-classical nature of the quantum state that is formed using QND measurements.

	\section{Summary and Conclusions \label{sec:Discussion}}

We have developed a theory of QND measurements for a single atomic ensemble.  The main result of this paper is the POVM (\ref{eq:measurementoperator_spectral}) which summarizes the effect of performing a QND measurement.  The basic effect of the measurement is described by the amplitude function (\ref{afuncdef}), which modulates the initial atomic wavefunction.  For small interaction times, the amplitude function takes the form of a Gaussian, with a width that decreases with both photon number and interaction time.  The position of the Gaussian depends upon the normalized difference of the photon counts according to relation (\ref{peakposition}).  For longer interaction times, the amplitude function develops non-Gaussian characteristics, which can be taken advantage of to create more exotic non-classical states such as Schrodinger cat states.  We also showed how in the large photon number limit the QND measurement reduces to a projective measurement onto the state $ | J, m_z \rangle $.  

Current experiments typically work in the regime of short interaction times, hence the QND measurement should be well-approximated by  the Gaussian form 
 (\ref{eq:measurementoperator_approx}).  As seen in Fig. \ref{fig3}, even after completely collapsing the state to $ | J, m_z \rangle $, photonic shot noise prevents one from precisely determining the true value of $ m_z $ for a single shot measurement.  Thus to determine $ m_z $, one would need to use a large photon number to reduce shot noise, make a large number of repeated measurements, or both.  One way to evade this would be to tune the phase $ \phi_{\chi \gamma}$ such that the photon probability emerges entirely at $ n_c = 0 $ (or equivalently $ n_d = 0 $).  In this way, effectively the photon number distribution is squeezed and one obtains a more accurate measurement of $ m_z $ from (\ref{peakposition}).  The precise understanding of QND measurements may allow for novel applications beyond squeezing and precision measurements~\cite{ilo-okeke2022,mao2022measurement,kondappan2022imaginary}.  One example is in realizing imaginary time evolution, where it was shown that combining QND measurements with a adaptive unitary can deterministically prepare eigenstates of various Hamiltonians, including cluster states \cite{mao2022measurement,kondappan2022imaginary}.

\begin{acknowledgments}
This work is supported by the National Natural Science Foundation of China (62071301); NYU-ECNU Institute of Physics at NYU Shanghai; the Joint Physics Research Institute Challenge Grant; the Science and Technology Commission of Shanghai Municipality (19XD1423000,22ZR1444600); the NYU Shanghai Boost Fund; the China Foreign Experts Program (G2021013002L); the NYU Shanghai Major-Grants Seed Fund; Tamkeen under the NYU Abu Dhabi Research Institute grant CG008. E.O.I.O. and B. C. A. acknowledge the support of the Tertiary Education Trust fund (TETFUND), Nigeria.
	\end{acknowledgments}

\section*{AUTHOR DECLARATIONS}
\subsection*{Conflict of Interest}
The authors declare no conflict of interest.

\section*{Data availability}
The data that support the findings of this study are available from the corresponding author upon reasonable request.

\appendix

	\bibliographystyle{apsrev4-1}
	\bibliography{ReferenceFile}

	%
	%
\end{document}